\RequirePackage{lineno}
\documentclass[reprint,longbibliography]{revtex4-2} %this defines a document style, in this case REVTEX is mostly for APS as PRL, but it is mostly a standard for pre-prints and ok in most journals
\usepackage{graphicx}%allows you to load figures
\usepackage{dcolumn}% allows two columns
\usepackage{bm}% use of bold fonts
\usepackage{color}% for color text ofc
\usepackage{caption}% for figure captions in the SI
\usepackage{amsmath}
\usepackage{caption} %correcting the FIG to figure
\usepackage{amssymb} %symbol of approximation
\usepackage[dvipsnames]{xcolor} %textcolor
\usepackage{ulem} % strikethrough text
\usepackage{soul} %underline text
\pdfoutput=1
\usepackage[T1]{fontenc} % something Oleg needed
\usepackage[utf8x]{inputenc}% something Oleg

\makeatletter %starting the first page%
\newcommand*{\balancecolsandclearpage}{%
  \close@column@grid
  \cleardoublepage
  \twocolumngrid
}

\makeatother

\begin{document}
\preprint{}
%\linenumbers

\title{Water Induced Ferroelectric Switching: The Crucial Role of Collective Dynamics}
%\textcolor{green}{A water switch for ferroelectric graphene edges}
%Layer-Dependent Self-Stabilizing Ferroelectric Behaviour of Water Clusters on the Edges of Graphene Nanoribbons

\author{Muhammad Awais Aslam{$^{1}$}}\email{muhammad.aslam@unileoben.ac.at}
\author{Igor Stankovic{$^{2}$}}\email{igor@ipb.ac.rs}
\author{Gennadiy Murastov$^{1}$}
\author{Amy Carl{$^{3}$}}
\author{Zehao Song{$^{4}$}}
\author{Kenji Watanabe$^{5}$}
\author{Takashi Taniguchi$^{6}$}
\author{Alois Lugstein$^{4}$}
\author{Christian Teichert$^{1}$}
\author{Roman Gorbachev{$^{3}$}}
\author{Raul D. Rodriguez$^{7}$}
\author{Aleksandar Matkovi\'{c}$^{1}$}\email{aleksandar.matkovic@unileoben.ac.at}

\affiliation{$^1$ Chair of Physics, Department Physics, Mechanical Engineering, and Electrical Engineering, Montanuniversit\"at Leoben, Franz Josef Strasse 18, 8700 Leoben, Austria}
\affiliation{$^2$ Scientific Computing Laboratory, Center for the Study of Complex Systems, Institute of Physics Belgrade, University of Belgrade, 11080 Belgrade, Serbia}
\affiliation{$^3$ Department of Physics and Astronomy and National Graphene Institute, University of Manchester, Manchester, United Kingdom}
\affiliation{$^4$ Institute of Solid State Electronics, TU Wien, Gußhausstraße 25-25a, 1040 Vienna}
\affiliation{$^5$ Research Center for Electronic and Optical Materials, National Institute for Materials Science, 1- 1 Namiki, Tsukuba 305-0044, Japan}
\affiliation{$^6$ Research Center for Materials Nanoarchitectonics, National Institute for Materials Science, 1-1 Namiki, Tsukuba, 305-0044, Japan}
\affiliation{$^7$ Tomsk Polytechnic University, Lenina ave. 30, 634034, Tomsk, Russia}

\date{\today}

\begin{abstract}
% First three phases should be why it is of general interest. Why it is important for hot application? Why it is new?
\textbf{The interaction mechanisms of water with nanoscale geometries remain poorly understood. This study focuses on behaviour of water clusters under varying external electric fields with a particular focus on molecular ferroelectric devices. We employ a two-fold approach, combining experiments with large-scale molecular dynamics simulations on graphene nanoribbon field effect transistors. We show that bilayer graphene nanoribbons provide stable anchoring of water clusters on the oxygenated edges, resulting in a ferroelectric effect. A molecular dynamics model is then used to investigate water cluster behaviour under varying external electric fields. Finally, we show that these nanoribbons exhibit significant and persistent remanent fields that can be employed in ferroelectric heterostructures and neuromorphic circuits.}

\end{abstract}

\maketitle %two coloums start

%\balancecolsandclearpage
%\section{\label{sec:level1}}
%---INTRODUCTION--------------------------------

\textbf{Introduction}

% INTRO - general part
Despite being at the centre of human life, water molecules and their interaction dynamics with various nanoscaled media still remain elusive and fascinating \cite{garcia2022interfacial, marechal2006hydrogen}.~This is partly due to the diversity of ways in which water molecules interact with other materials. The prospect of affordable and efficient catalysts triggered studies investigating the relationship of solid surfaces especially that of graphene and other 2D materials with aqueous (H$_2$O) media \cite{su2018role, melios2018water, hong2014thermal, acik2011nature}. Moreover, understanding of water interaction with low dimensional materials has proven essential in the fields of nanofluidics \cite{emmerich2022enhanced}, energy storage \cite{mendoza2016synthesis}, water splitting \cite{cai2022wien}, purification \cite{dervin20162d,surwade2015water} and  water-assisted ferroelectricity \cite{chin2021ferroelectric, aslam2022single}. Especially, in nano-structured materials water can play a critical role in tuning the properties \cite{wang2011potential, kvashnin2013impact, jiang2018study, tan2013atomically, wagner2013band} and in ambient conditions its effects are often unavoidable.

% focus on why graphene nanoribbons, and finish with application spectrum
In this regard, a graphene nanoribbon with a high edge-to-surface ratio presents an excellent platform to study the influence of water behaviour at the edges. Despite extensive progress in the synthesis of graphene nanoribbons \cite{jiao2010facile, cai2010atomically},  water-terminated graphene edges have only been realized recently in a stable manner \cite{aslam2022single, caridad2018graphene}. The polar water molecules attached on the edges are functional features of the obtained field effect transistors exhibiting ferroelectric behaviour~\cite{aslam2022single, caridad2018graphene}.~Still, field-related kinesis of edge-adsorbed water, the influence of graphene thickness on water anchoring, and consequently their interplay on ferroelectricity are not fully uncovered yet.~An understanding of this surprisingly stable phenomenon is essential as it warrants the use of water-induced ferroelectricity in radio frequency applications, neuromorphic computing and memcapacitors \cite{liu2012ferroelectric, liu2020two, guan2020recent, li2021enhanced, krems2010ionic}.

% Paper overview (in as short as possible form)
In this work, we employ networks of graphene nanoribbons (Gr-NRs) integrated into field effect transistors (FETs) to demonstrate that at least bilayer thickness is required for a temperature-stable ferroelectric effect. The observed dependence on the number of layers, temperature, and the applied external electric fields is captured by our molecular dynamic (MD) model. We propose that the system exhibits collective behavior of water based on anchoring of the bridging water molecules between the layers. Therefore, the kinesis of the system is strikingly different in mono- and multi-layer nanoribbons. Lastly, we confirm the presence of the remanent field, as predicted by the MD simulations.

\begin{figure*}
\centering
\includegraphics[width=1\textwidth]{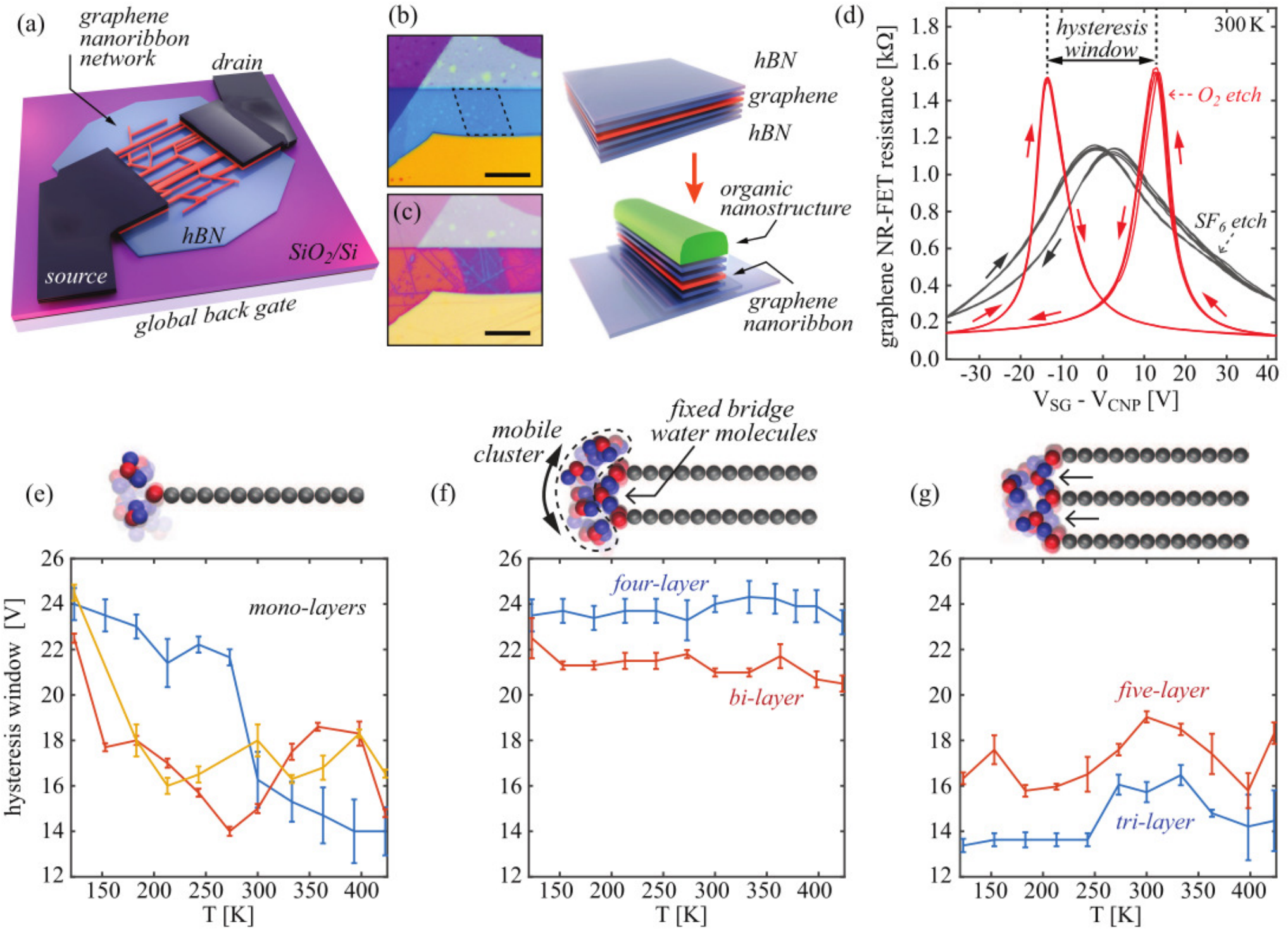}
\caption{Layer dependent stability of ferroelectricity in hBN encapsulated graphene nanoribbons. (a) Schematic representation of nanoribbon filed effect transistor. (b,c) Optical images (scale-bar 10~$\mu$m) and corresponding sketches for fabrication of hBN-Gr-hBN nanoribbons. (d) Transport characteristics for 2L Gr-NR-FETs etched using different precursor gases (SF$_6$ and O$_2$); in both cases presenting five subsequent sweeps at room temperature and under low vacuum. The arrows in (d) indicate the sweep direction. V$_{\rm SG}$ range is shifted with respect to the mean V$_{\rm CNP}$ values between both sweeping directions. (e-g) Temperature dependence of the hysteresis window considering varying thicknesses of graphene NRs. The plot for monolayers includes three different devices colored blue, yellow and orange. Top of (e-g) MD simulation models exhibiting typical local configuration of water clusters for at the Gr edges for varying thicknesses of graphene in absence of the external electric field. The colours represent type of atom: hydrogen (blue), oxygen (red) and carbon (grey).}
\label{Figure:1}
\end{figure*}

$ $\par
%\textbf{2 Results and discussions}
\textbf{Bridging mechanism for H$_2$O on graphene nanoribbons edges}

% This explains Fig1 a-d
Figure 1a depicts the schematic representation of a Gr-NR FET fabricated on hexagonal boron nitride (hBN). See supplementary Material section S1 for more information. Parahexaphenyl (6P) organic nanoneedles self-assembled on hBN-Gr-hBN heterostructures serve as self-aligned masks \cite{matkovic2019light, aslam2022single, kratzer2015effects}. Reactive ion etching of the stacks produces a network of Gr nanoribbons (Figure 1b,c). The choice of precursor gas for plasma controls nanoribbon termination. Using oxygen (O$_2$) or sulfur hexafluoride (SF$_6$) allows the edges to be either oxygen or fluorine terminated, respectively \cite{caridad2018graphene}. Our devices with oxygen-terminated edges exhibits symmetric, pronounced, stable, and switchable (between $p-$ and $n-$doping) characteristics depending on the global back gate (V$_{\rm SG}$) sweeping direction (Figure~1d). The comparison between the nanoribbons and flake FETs is represented in Sec. S2 of the Supplemental Material.  The difference between two charge neutrality points (CNPs - $R(V_{\rm SG})$ maxima in either positive or negative sweeping direction) is quantified as the hysteresis window of the ferroelectric nanoribbons. The relationship between the direction of the applied field and the resulting gate bias where the resistivity maxima occurs (V$_{\rm CNP}$) have been explained in more detail in the Sec. S3 of the Supplemental Material. The hysteretic behaviour is almost absent in hydrophobic SF$_{6}$ etched \textit{i.e.} $F$-terminated ribbons, as has been previously observed for graphene flakes \cite{caridad2018graphene}. From this point onwards we will only discuss oxygen-terminated graphene nanoribbons, which support the adsorption of water molecules and the induced ferroelectricity \cite{berashevich2010doping, lin2013adsorption, caridad2018graphene}.

% This introduces the hysteresis origin and previous observations
The contours of the mechanism behind the hysteretic characteristics shown in Figure~1(d) are known, as Caridad $et.al.$ proposed that single molecules switch between the two states~\cite{caridad2018graphene}. The external gate fields disrupt the equal probability of water molecule arrangements while inducing a torque on the water adsorbed at the edges. The torque is a result of Coulomb forces acting on the positively charged hydrogen atoms and negatively charged oxygen atoms, the energy barrier between the two states is $ca.$ 25~-~40~meV~\cite{yamamoto2017water,caridad2018graphene}.

% HERE first explain the observations
Strikingly, the $R(V_{\rm SG})$ hysteresis is thermally stable while at the same time it depends on the number of graphene layers. Figure~1e-g compares monolayer (ML) to thicker Gr-NRs over a temperature range of 120~K~-~400~K. In monolayer Gr-NR FETs, the width of the hysteresis window decreases as temperature increases. However, even within the temperature range tested, the hysteresis window remains significant. The width of hysteresis for L$\geq$2 (L~-~number of graphene layers) is independent of temperature (Figure~1f,g). Compared to monolayer, the temperature characteristics of three- and five-layered graphene show less stable behavior. However, the hysteresis window does not exhibit a decreasing trend. This suggests a mechanism involving a single water molecule crossing the barrier and switching the side of the graphene plane, which would result in a stronger temperature dependence. When the thermal energy of the molecule is larger than the energy barrier, molecules can thermally switch from one state to the other. Such behaviour was observed in the case of monolayer Gr-NR FETs, with the hysteresis still preserved at elevated temperatures (Figure~1e). However, the results for thick (L$\geq$2) NR FETs indicate that the mechanism involved in the observed ferroelectric effect is not thermally activated (Figure~1f,g).

%Proposing conditions that are needed to observe such behaviour
The tendency of the water molecules to bind at oxygenated edges and form clusters \cite{offei2023collective} suggests that collective behaviour stabilizes the molecules resulting in a temperature independent effect. To achieve such stable ferroelectric ordering of water molecules on Gr edges, two conditions must be met: $(i)$ a low thermal barrier of a single polar molecule to switch from one state to the other, and $(ii)$ a cluster large enough to stay bound in one state by intermolecular Coulomb interactions. While the first condition is well established, we have designed molecular dynamics simulations to investigate the second condition.

% Now hit the readers with the MD simulations
Figures~1e-g show cross-sectional side-views from a molecular dynamics (MD) simulation without an external field, depicting the water molecules anchoring to the sides of the ribbons. Transparency of the H$_2$O molecules in the figure indicates the distance from the cross-section plane. The MD simulations of single-layer graphene confirm the previous prediction of a 25~meV-40~meV energy barrier for the switching of a single water molecule~\cite{yamamoto2017water,caridad2018graphene}. For the bi-layer system (top of Fig.~1f), the simulations reveal a fundamentally different behaviour of the water molecules. The particular layer separation in graphite allows water molecules to form a "bridge" between the two oxygen-terminated edges. These bridging H$_2$O molecules remain stable and do not react to applied external electric fields. However, they enable the formation of a H$_2$O cluster around them. The strong Coulomb interaction of positively charged hydrogens in water, anchor the water molecules to oxygen atoms in the graphene edge and immobilise the bridge water. Molecular dynamics simulations show that anchored water molecules in the cluster can sustain both high electric fields (up to 5~V/nm) and elevated temperatures up to 500~K. Still, the water cluster that surrounds the bridge molecules is mobile in the electric field, and its collective behaviour stabilizes the structure, as a result the hysteresis window is essentially temperature-independent. Such collective self-stabilizing behaviour of polar objects was previously observed in colloidal systems and molecular motors \cite{KLAPP2016,Lipowsky2005}.

\begin{figure*}
\centering
\includegraphics[width=1\textwidth]{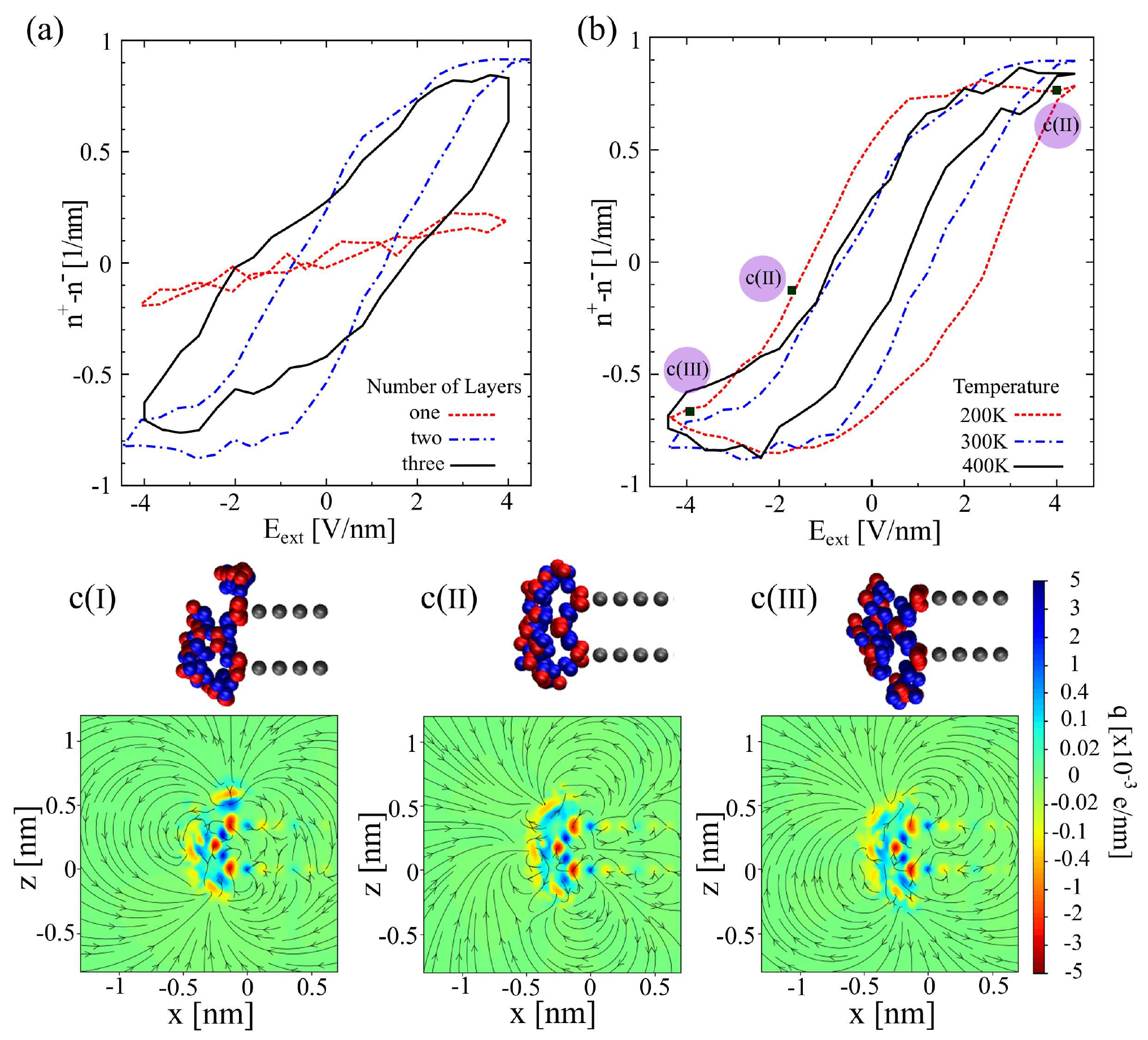}
\caption{ Evolution of two populations of water molecules “above” $n^{+}$ and “below” $n^{-}$ the graphene planes with the cyclical change of homogeneous external electric field between $E_{\rm ext}$=-4.1 and 4.1~V/nm. (a) With varying thicknesses of graphene. (b) with varying temperature. (c) The panels (I - III) illustrate induced electric fields $E$ of a graphene ribbon and water molecules at (I) $E_{\rm ext}=$4~V/nm, (II) $E_{\rm ext}=$-2~V/nm, and  (III) $E_{\rm ext}=$-4~V/nm.  Respective density of charges as well as electric field lines are shown in c(I), c(II), and c(III). The color represents the type of atom: hydrogen (blue), oxygen (red), and carbon (grey). Simulated is a bilayer graphene ribbon (periodic in the y direction) of width $W$=~23~nm. }
\label{Figure:2}
\end{figure*}
$ $\par

\textbf{Water clusters under electric fields}
$ $\par
We performed MD simulations to understand the effect of varying electric fields that act on H$_2$O molecules. An external homogeneous electric field generated by infinite planes was applied to ML, 2L and 3L graphene NRs. Refer to the Supplementary Material S1 for details about MD simulations. The $n^{+}$ and $n^{-}$ are linear densities of H$_2$O molecules adsorbed to the graphene edge above and below the graphene layer's center of mass, respectively. The fraction $n^{+}-n^{-}$ of polarised molecules above ($n^{+}$) and below ($n^{-}$) the middle plane of the graphene layer quantifies the ferroelectric effect arising from molecular switching. We observed charge bistability and switching between two states under cyclical change of the external field, as indicated by the experimental results. This is represented in Figure 2a.

For a monolayer ribbon, the fraction of molecules ($n^{+}-n^{-}$) is proportional to the external electric field, suggesting a low energy barrier between the states. This can be attributed to the absence of a bridging H$_2$O molecules compared to systems with L~$\geq$~2. However, our experiments on monolayer graphene nanoribbons showed a temperature-dependent hysteresis, which may be attributed to the larger ribbon system and more complex interfaces available experimentally. Additionally, the shorter timescale of the molecular dynamics simulations could be a factor. The edges of the hBN ribbons which encapsulate graphene and the remaining organic mask could also support the H$_2$O cluster formation, however the system is more complex and less well-defined compared to bilayer graphene nanoribbons.

For L $\geq$ 2, a prominent evolution of the hysteresis was noted in the MD simulations as shown in Figure~2a. Such an evolution of the system with electric field depends on the initial state. The field lines and field strength acting on a 2L nanoribbon generated via MD simulations are represented in Sec. S4 of the Supplemental Material. As the electric field decreases, some of the molecules remain stationary until the field polarization switches, causing these molecules to migrate to the other side with respect to the nanoribbon's basal plane. It should be noted that the hysteresis is weakly dependent on the temperature as shown in Figure~2b for L~=~2. Such stability indicates that the switching of polarization is also not thermally activated in the model as increased thermal energy (400~K) does not result in disorder that could change the ferroelectric nature of the system.

\begin{figure*}
\centering
\includegraphics[width=1\textwidth]{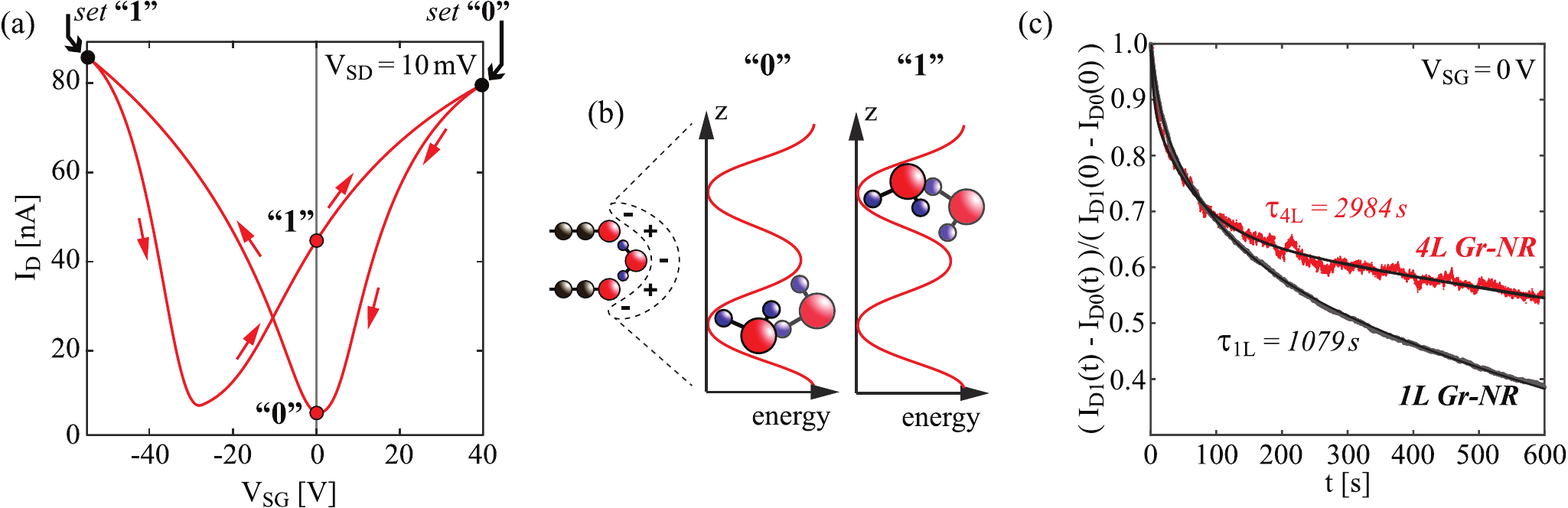}
\caption{Field remanence after bias stressing. (a) Shows the hysteresis curve for Gr-NR FET after the application of a bias stress. (b) The schematic representation of retention states for the water clusters adsorbed at the edges. (c) The source-drain current deviation in time with respect to the ML vs 4L graphene nanoribbon devices. The time constant was evaluated by fitting the curve with exponential decay for each state ("1" and "0"). All measurements were done at room temperature.}
\label{Figure: 3}
\end{figure*}

Figure 2c (I-III) represents snapshots of the edge segments for a 2L Gr-NR system with the corresponding charge distribution and induced electric filed lines. The system consists of both externally applied fields (not shown in Figure~2c) and induced fields with opposite polarities in \textbf{z} direction. Figure 2c(I) presents the case of saturation with an external field of 4.1~V/nm applied in \textbf{z} direction. An induced charge redistribution is observed due to the molecular polarisation at the edges, which in turn creates an effective electric dipole moment. This changes the local field over the whole graphene double-layer profile. The effective electric field is, therefore, a sum of the two fields and as the external field decreases, they could cancel each other in parts of the graphene plane. Figure 2c(II) displays the configuration where the induced electric dipole points roughly along graphene, $i.e.$ representing one of the coercivity points of the out-of-plane dipole field hysteresis. Consequently, the induced electric field component perpendicular to the ribbon will be negligible. Finally, for -4.1~V/nm we observe a complete reversal of the induced electric field. As the field strength increases, the migration and molecular polarisation repeat in the opposite order, generating a hysteresis loop (see Figure~2b). Furthermore, the bonding between the polar molecular ensemble and the graphene edge, together with the intermolecular Coulomb interactions should be strong enough to prevent the external electric field from tearing off molecules from the cluster. We designed a large-scale atomistic model in order to validate the proposed mechanism of collective H$_2$O molecule switching which is represented in Sec. S5 of the Supplemental Material. Our MD simulations show that under external fields exceeding $\sim$5~V/nm, individual molecules dissociate from the edges at 300~K.

$ $\par
\textbf{Remanence of the net induced dipolar field}
$ $\par
A pronounced hysteresis in the electrical transport curves of Gr-NR-FETs terminated with H$_2$O molecules does not indicate the remanence of their net dipolar field. To confirm the pronounced hysteresis predicted by the MD simulation, we tracked the evolution of the drain current (I$\rm _D$) of Gr NR FETs without an external gate field, after pre-biasing the devices into the $n^{+}$ or the $n^{-}$ states. Figure~3a presents a hysteretic transfer curve of the H$_2$O terminated Gr-NR-FET, where bias stress via an asymmetric V$\rm _{SG}$ sweeping was done to position one of the CNPs near V$\rm _{SG}$~=~0~V~\cite{bagheri2021control}. Sec. S6 of the Supplemental Material represents the sweeping cycles repeated symmetrically and asymmetrically. This way the difference in I$\rm _D$ for the $n^{+}$ and the $n^{-}$ states is maximized. The two I$\rm _D$(V$\rm _{SG}$~=~0~V) states can be defined as "0" and "1", corresponding to low or the CNP state and the high current state. Figure~3b depicts a sketch, based on the MD simulations, of the two retentivity states of the edge adsorbed water clusters in the case of a 2L Gr-NR, relating the states of the dipoles to the "0" and "1" states of the I$\rm _{D}$. The tested Gr-NR-FETs exhibited one order of magnitude difference in current between the two states.

By sweeping from V$\rm _{SG}$~=~0~V to one end of the range, and returning back to V$\rm _{SG}$~=~0~V, the system is set either to the I$\rm _{D}$ "0" or "1" state. From this point on, the I$\rm _{D}$ is recorded as a function of time without an externally applied gate field. In both cases of mono- and multi-layer nanoribbons, the dipole induced remanent field was observed. The results comparing a ML and a 4L Gr NR FET are shown in Figure~3c, where the normalized difference between I$\rm _{D1}$(t) and I$\rm _{D0}$(t) was tracked for 600~s at 300~K and under low vacuum. To characterize the decay of the field, the curves were fitted to $C\cdot exp(-t/\tau)$; where $C$ stands for a scaling constant and $\tau$ for time constant.

As was also observed in the temperature dependence of the hysteresis window, the remanence of the field for a ML is less pronounced and the current difference between the two states decays faster than in the multi-layer Gr NR FETs, as shown in Figure 3c. For ML Gr NR FETs $\tau$~=~(1041~$\pm$~315)~s was obtained, as a mean value considering multiple measurements from several devices. In contrast, 4L NRs showed about three times longer retention ($\tau$~=~(3020~$\pm$~125)~s). The time constants indicate that the ML NR-based devices lose 90~\% of the initial  I$\rm _D$ difference between the "0" and "1" states in about 30~min, and 4L NR-based devices in about 90~min. Further, a faster decaying component ($\tau$~=~(64~$\pm$~32)~s) responsible for about 15-20~\% of the I$\rm _D$ difference was noticed in all devices, and can be attributed to molecules that are weakly bound in the cluster. Gr-NR-FETs with L~$\geq$~2, require several hours of storage in ambient and room temperature to have the initial transfer curve sweep starting from the depolarized state of the water molecules (as shown in Fig.~S2), which is consistent with the extracted time constants.

\textbf{Summary}\par
% what we have observed and what it could be
In summary, hBN-encapsulated graphene nanoribbon FETs were employed to study the temperature and thickness dependent stability of water induced ferroelectricity. A different trend in the temperature dependency of the hysteresis window was observed for monolayer than for L~$\geq$~2 nanoribbons. In multilayer NRs the switching of polarization is not thermally activated, which indicates a collective and self-stabilizing effect. This behaviour would require a H$_2$O cluster large enough so that collectively molecules can stay bound in one state by intermolecular Coulomb interactions.

% what we think it is based on the model
MD simulations were used to propose a mechanism of water kinesis at the edges of nanoribbons that explains the temperature and thickness dependence of the induced dipole fields observed experimentally. The simulations revealed a fundamentally different behaviour of the water molecules for mono- and multi-layer NRs. For L~$\geq$~2, the two adjacent layers enabled the adsorption of a "bridge" H$_2$O molecule, that together with the oxygen-terminated edges promotes the formation of a water cluster. Our model predicts collective behavior of the H$_2$O cluster and bi-stability in the external electric fields, resulting in hysteresis of the induced dipole fields acting on the Gr-NRs.

% finishing flavors
We confirmed the existence of the remanent water dipole-induced fields, and their temporal evolution. These robust, temperature independent, and high remanent fields in graphene nanoribbon networks could be utilized in the heterostacks with other 2D materials, creating new device concept based on molecular switching for applications such as computing in memory, and synaptic circuits. Our work constitutes as a nascent effort to comprehend the interaction dynamics between water molecules and the periphery of nanostructured materials.

$ $\par

\par$ $\par
%\bibliographystyle{plain}
%\bibliography{REFERENCES}

\par
$ $\par
$ $\par

\par$ $\par
{\textbf{Acknowledgements}}{This work is supported by the Austrian Science Fund (FWF) under grants no. I4323-N36 and Y1298-N, and by the Russian foundation for basic research under the project no. 19-52-14006. I.S. acknowledges the support of the Ministry of Education, Science, and Technological Development of the Republic of Serbia through the Institute of Physics Belgrade and the European Commission through the ULTIMATE-I project partner Senzor Infiz doo, grant ID 101007825. Molecular dynamics calculations were run on the PARADOX super-computing facility at the Scientific Computing Laboratory, Center for the Study of Complex Systems of the Institute of Physics Belgrade. K.W. and T.T. acknowledge support from the JSPS KAKENHI (Grant Numbers 20H00354 and 23H02052) and World Premier International Research Center Initiative (WPI), MEXT, Japan. R.G. acknowledges support from Royal Society, ERC Consolidator grant QTWIST (101001515) and EPSRC grant numbers EP/V007033/1, EP/S030719/1 and EP/V026496/1. A.C. acknowledges support from EPSRC CDT Graphene NOWNANO, grant EP/L01548X/1.
\par $ $ \par

{\textbf{Author Contributions }}{MAA prepared the samples, carried out experiments and data analysis under the supervision of AM. IS performed the simulations. GM performed the measurements for the field remanence. MAA wrote the manuscript with the help of IS and AM. MAA and ZS carried out etching experiments under the supervision of AL. KW and TT provided hexagonal boron nitride crystals. RG and AC provided monolayer graphene on hexagonal boron nitride. RDR and CT helped in the internal review of the manuscript. AM and RDR acquired the main funding for the study. All the authors discussed the results and reviewed the manuscript. }

\par$ $\par
%\par$ $\par
{ \textbf{Additional information}}\par
{\scriptsize \textbf{Supporting Information:} The online version contains the Supplemental Material.}\par

{\scriptsize \textbf{Competing financial interests:} The authors declare no competing financial interests.}\par

{\scriptsize \textbf{Keywords:} Water, Edges, Ferroelectricity, Nanoribbons, Graphene}\par

\end{document}